\documentclass[
  twocolumn,
  superscriptaddress,
  amsmath,amssymb,
  aps,
  pra,
  floatfix,
10 pt]{revtex4-2}

\usepackage{graphicx}    
\usepackage{tikz} 
\usepackage{dcolumn}     
\usepackage{bm}  
\usepackage{placeins}
\usepackage{xcolor}
\usepackage{physics}     
\usepackage{xcolor}      
\usepackage{soul}        
\usepackage{mleftright}  
\usepackage[colorlinks]{hyperref}
\usepackage[capitalise]{cleveref}
\hypersetup{
    colorlinks = true,
    linkcolor  = blue,
    citecolor  = blue,
    urlcolor   = blue
}

\definecolor{navyblue}{rgb}{0.0, 0.0, 0.5}

\begin{document}

 \title{Reservoir-induced Quantum Phase Transitions in spin ensembles}
\author{Giuseppe Musolino}
\affiliation{Dipartimento di Scienze Matematiche e Informatiche, Scienze Fisiche e Scienze della Terra, Universit\`{a} di Messina, I-98166 Messina, Italy}

\author{Daniele Lamberto}
\email{daniele.lamberto@riken.jp}
\affiliation{Quantum Information Physics Theory Research Team, RIKEN Center for Quantum Computing, Wakoshi, Saitama, 351-0198, Japan}
\affiliation{Dipartimento di Scienze Matematiche e Informatiche, Scienze Fisiche e Scienze della Terra, Universit\`{a} di Messina, I-98166 Messina, Italy}

\author{Gabriele Orlando}
\affiliation{Dipartimento di Scienze Matematiche e Informatiche, Scienze Fisiche e Scienze della Terra, Universit\`{a} di Messina, I-98166 Messina, Italy}

\author{Omar Di Stefano}
\affiliation{Dipartimento di Scienze Matematiche e Informatiche, Scienze Fisiche e Scienze della Terra, Universit\`{a} di Messina, I-98166 Messina, Italy}

\author{Salvatore Savasta}
\affiliation{Dipartimento di Scienze Matematiche e Informatiche, Scienze Fisiche e Scienze della Terra, Universit\`{a} di Messina, I-98166 Messina, Italy}

\begin{abstract}
In this work, we present a comprehensive analysis of an open quantum system consisting of a collection of $N$ two-level atoms in the thermodynamic limit coupled to bosonic reservoirs.
Our results demonstrate that, when the system-bath interactions lacks a metastable minimum, as in the counterterm-free case, the system can undergo a quantum phase transition set by the reservoir structure rather than by coherent internal dynamics.
Using exact quantum Langevin equations, valid even in the overdamped regime near criticality, where the loss rates can exceed the resonance frequency of the softening mode, we locate the critical point via the complex excitation spectrum and obtain closed-form reflection and transmission spectra that display the mode softening and an increasingly asymmetric lineshape near the transition.
\end{abstract}

\maketitle

\section{Introduction}
\label{sec:intro}

A deeper perspective on open quantum systems \cite{breuer2002theory,RevModPhys.88.021002,PhysRevLett.110.130406,PhysRevLett.109.170402}, now well established~\cite{Diehl2008,PhysRevLett.110.195301}, recognizes dissipation as an active structural ingredient capable of inducing macroscopic quantum phenomena \cite{walls2008quantum,9zl7-31f3}, including quantum phase transitions (QPTs) \cite{PhysRevLett.119.053601,Sachdev_2011,Kim2025,Chen2024,Osterloh2002} driven by system-bath interactions rather than by coherent internal dynamics \cite{Senthil2004}.
Such reservoir-induced QPTs include bosonic baths driving the localization-delocalization transition of the spin-boson model \cite{RevModPhys.59.1}, the interplay between multiple baths in multi-bath spin-boson models \cite{PhysRevB.90.245130}, and dissipation-induced transitions in the quantum Rabi model \cite{PhysRevLett.130.210404,PhysRevA.97.013825}.
Identifying and characterizing such transitions \cite{DiBello2024,Lin2016} is both a fundamental question in the theory of open quantum systems and an opportunity for engineering quantum criticality \cite{PhysRevLett.126.010502,PRXQuantum.3.010354,PhysRevLett.132.060801,cxvs-5pb1,RevModPhys.89.035002,PhysRevLett.117.110802,PhysRevLett.130.123601,PhysRevLett.124.120504,PRXQuantum.6.020301}.
However, the mechanism studied here, in which a counterterm-free reservoir renormalizes the system frequency to zero, has so far been confined to classical and single-particle settings.

Describing these systems near a critical point poses a fundamental challenge \cite{Kobayashi2025,Kremen2018,PhysRevA.84.043637}. As the lowest eigenfrequency vanishes upon mode softening, standard approximations that assume loss rates much smaller than the relevant system frequencies (e.g., Born-Markov and rotating-wave \cite{PhysRevA.80.033846}) fail, and the system inevitably enters the ultrastrong system-bath coupling regime \cite{FriskKockum2019,PhysRevA.98.053834,PhysRevLett.130.123601,PhysRevLett.112.016401}.
Quantum Langevin equations (QLEs) overcome this limitation: unlike master-equation approaches \cite{PhysRevA.87.043840,PhysRevResearch.6.033181}, they retain the full counter-rotating structure of the system-bath interaction and yield exact equations of motion valid for arbitrary coupling strengths and bath spectral densities.

The microscopic structure of the system-bath interaction determines whether the coupling stabilizes or destabilizes the system.
Within the Caldeira-Leggett framework~\cite{PhysRevLett.46.211,Kovacs2017}, the standard bilinear-plus-counterterm coupling generates a displaced harmonic potential for each bath mode, preserving the stability of the total Hamiltonian \cite{CALDEIRA1983374}.
The purely bilinear interaction, by contrast, lacks any stabilizing minimum and renormalizes the system frequency downward by an amount proportional to the integrated coupling strength $\chi$.
When $\chi$ reaches a critical value $\chi_c$, the renormalized frequency vanishes, signaling a reservoir-induced QPT \cite{Sachdev_2011}; when two such channels compete, a rich phase diagram of dissipative transitions can emerge \cite{PhysRevB.90.245130}.
This mechanism, known in the Caldeira-Leggett literature, has not previously been exploited as a driver of many-body QPTs in ensembles of two-level systems (TLSs) \cite{PhysRev.93.99}, nor connected to experimentally accessible spectroscopic signatures.

This route to criticality is conceptually distinct from the Dicke superradiant phase transition (SPT), which it superficially resembles.
In the SPT, recently observed on a magnonic platform \cite{Kim2025}, mode softening and macroscopic condensation are driven by the \emph{coherent} light-matter coupling exceeding a threshold, while the diamagnetic $A^2$ term tends to forbid the transition \cite{Nataf2010,PhysRevB.100.121109,PhysRevResearch.7.013271}.
Here, by contrast, no coherent drive or cavity mode is involved: the transition is driven by the \emph{dissipative} reservoir alone.
Softening and condensation thus arise from the equilibrium structure of the system-bath coupling itself.

In this work, we study a minimal model of $N$ two-level systems (TLSs) \cite{Santanni2024Metalloporphyrins,10.1021/jacs.8b06733,10.1039/c6cp08161d,Bader2014} in the thermodynamic limit ($N \to \infty$) \cite{PhysRevLett.92.073602}, simultaneously coupled to two structurally distinct bosonic reservoirs \cite{PhysRevB.104.L060410,DiBello2023}: the first through a displaced harmonic potential with a metastable energy minimum, the second through a purely bilinear interaction without stabilizing structure. Combining the Holstein-Primakoff mapping with QLEs, we derive the exact dynamics on both sides of the critical point $\chi_c$, without resorting to perturbative or Markovian approximations \cite{Merkli2022dynamicsofopen,PhysRevLett.115.043601,PhysRevA.97.052116}. 
We show that the critical point is determined by the simultaneous vanishing of both the real and imaginary parts of the complex excitation frequency, and that it is independent of the dissipation rates. In the abnormal phase ($\chi > \chi_c$), a global minimization of the total system-bath Hamiltonian reveals macroscopic ground-state occupation \cite{PhysRevB.100.121109,PhysRevLett.127.167201} not only in the system but simultaneously in both reservoirs, driven by the unstable bath.

From this framework we derive compact closed-form expressions for the reflection and transmission coefficients, $S_{11}(\omega)$ and $S_{21}(\omega)$, valid for any bath spectral densities and arbitrarily large damping rates. 
Near the critical point  \cite{Hayashida2023,PhysRevLett.127.207202}, the corresponding spectra exhibit two distinctive features: an asymmetric Fano-like lineshape \cite{Limonov2017,RevModPhys.82.2257}, originating from the counter-rotating structure of the system-bath coupling and becoming increasingly pronounced as $\chi \to \chi_c$; and the exact cancellation of the Lamb shift \cite{Silveri2019} in both phases, implying that fits based on closed-system eigenfrequencies remain accurate even deep in the critical region. 
As a concrete realization \cite{Ding2022}, we propose a yttrium iron garnet thin film coupled to an open superconducting coplanar waveguide, where the magnetic (Zeeman) \cite{PhysRevResearch.6.013303,PhysRevApplied.20.024039} and electric (diamagnetic) \cite{PhysRev.80.440} interactions naturally furnish the two structurally distinct reservoirs, and where the recent achievement of ultrastrong magnon-photon coupling brings this QPT within experimental reach \cite{PhysRevB.72.115303,Noel2022,PhysRevLett.132.176601,Greiner2002}.

\section{Model}
\label{sec:model}

We consider a system of $N$ identical, non-interacting TLSs of resonance frequency $\omega_a$,
\begin{equation}
\label{collection}
   H_{\rm sys}= \frac{\hbar \omega_{a}}{2}\sum_{k=1}^{N}\sigma^{z}_{k}\, ,
\end{equation}
where $\sigma_{k}^{z}$ is the Pauli operator associated to the $k$-th TLS. The corresponding total spin operator is $J^{z}=\hbar\sum_{k}\sigma_{k}^{z}/2$, which obeys the $SU(2)$ algebra \cite{sakurai2020modern}. Our analysis centers on the maximum-cooperativity subspace $j=N/2$, to which the ground state is confined at zero temperature in the large-$N$ limit \cite{Nataf2010,PhysRevE.67.066203,PhysRevLett.90.044101,rgz2-4m69}.

\begin{figure}[t]
    \centering
    \includegraphics[width=\columnwidth]{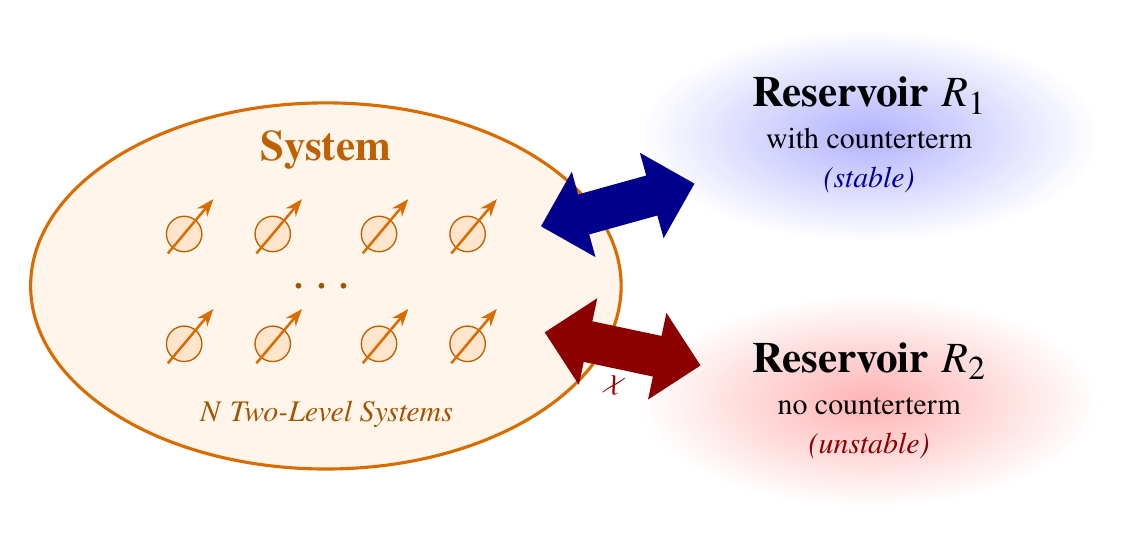}
    \caption{Open quantum system consisting of a collection of $N$ TLSs in the thermodynamic limit ($N\to \infty$) simultaneously coupled to two structurally distinct bosonic reservoirs. Reservoir $R_1$ couples through a displaced harmonic potential (bilinear plus counterterm), which features a stable energy minimum. Reservoir $R_2$ couples through a purely bilinear interaction that lacks any such stabilizing structure. Here, the integrated coupling strength $\chi$ quantifies the downward renormalization of the system frequency and drives a quantum phase transition when it reaches the critical value $\chi_c$.}
    \label{open system}
\end{figure}

As in the Caldeira-Leggett framework, each reservoir is modeled as an infinite set of uncoupled harmonic oscillators coupled to the system through the transverse-spin coordinate $X_A\propto J^x$. The first reservoir couples through a displaced harmonic potential (bilinear-plus-counterterm), with a metastable minimum; the second through a purely bilinear interaction with no stabilizing structure (\cref{open system}). The total Hamiltonian can be written as
\begin{equation}
\label{eq:renormalized}
    H = H'_{\mathrm{sys}} + \sum_{j=1,2} \sum_n \left( \frac{p_{jn}^2}{2} + \frac{k_{jn}}{2} (q_{jn} - X_A)^2 \right)\, ,
\end{equation}
where $k_{jn}$ are the microscopic system-bath couplings and $H'_{sys}$ is the renormalized system Hamiltonian 
\begin{equation}
\label{eq:Hprim}
    H'_{\mathrm{sys}} = H_{\mathrm{sys}} - \frac{\chi}{2}  X_A^2\, .
\end{equation}
Here $\chi=\sum_n k_{2n}$ is the integrated coupling strength of the second reservoir and it is related to the frequency integral of the reservoir decay rate $\gamma_2(\omega)$, $\chi= \int_0^\infty d\omega\, 2\gamma_2(\omega) /\pi$ (see \cref{Memory_f_decay}). For a frequency-independent damping rate $\gamma_2$ up to a cutoff $\omega_c$, this reduces to the familiar Caldeira-Leggett relation $\chi \propto \gamma_2\omega_c$ \cite{RevModPhys.59.1}.
$\chi$ is the key quantity driving the reservoir-induced QPT: dynamical dissipation on the system is set by $\gamma_2$ close to the system frequency, whereas the renormalization in \cref{eq:Hprim} is set by its integral $\chi$, and the two can be modelled independently. In particular, even when $\gamma_2(\omega)\approx0$ across the low-frequency range where the QPT resides (so that no bath mode exchanges energy with it resonantly and the dissipation vanishes), $\chi$ remains finite as its value is dominated by the far-detuned high-frequency continuum \cite{ripoll2022quantum}. These off-resonant modes cannot drive real transitions but dress the system statically, shifting its frequency in close analogy with a phononic Lamb shift \cite{PhysRevX.6.041041}. We refer to the limiting regime of $\chi$ finite while the dynamical dissipation is negligible as the \emph{pseudo-isolated} system, $H'_{\rm sys}$.

In the thermodynamic limit ($N\to\infty$ at fixed $\rho=N/V$), the Holstein-Primakoff mapping \cite{PhysRev.58.1098,grosso2013solid} gives $H_{\rm sys}=\hbar\omega_a a^\dagger a$ and $X_A=\sqrt{\hbar/2\omega_a}\,(a^\dagger\sqrt{1-a^\dagger a/N} + \sqrt{1-a^\dagger a/N}\,a)$. The nature of the ground state of the pseudo-isolated system distinguishes two regimes: in the \emph{normal phase} (NP), the mode is not macroscopically occupied, $\langle a^\dagger a\rangle/N = 0$, so the square roots reduce to unity and $X_A=\sqrt{1/2\omega_a}\,(a^\dagger+a)$, while in the \emph{abnormal phase} (AP) the mode acquires a macroscopic occupation, $\langle a^\dagger a\rangle/N \neq 0$. Restricting $H'_{\rm sys}$ to the NP defines the Hamiltonian $H'_{NP}$, whose positive eigenvalue represents the pseudo-isolated eigenfrequency
\begin{equation}
\label{omega0NP}
\tilde\omega_0=\omega_a\sqrt{1-\chi/\omega_a^2}\, ,
\end{equation}
which vanishes as $\chi\to \chi_c=\omega_a^2$, signaling the onset of the QPT and the entry into the AP.

In the AP, the macroscopic occupation is captured by displacing $a=a_s+\alpha$, with $\alpha$ a c-number of order $\mathcal{O}(N)$ and $a_s$ the fluctuation operator. Minimizing \cref{eq:renormalized}, we obtain (see SM)
\begin{equation}
\label{alpha}
    \alpha^2= \frac{N}{2}\left(1-\frac{1}{\eta}\right)\, ,\qquad
    \eta=\chi/\omega_a^2\, .
\end{equation}
This displacement yields the abnormal-phase Hamiltonian $H'_{AP}=\bar{\omega}_a a_s^{\dagger}a_s+(C-\chi/4\bar{\omega}_a \eta)(a_s^{\dagger}+a_s)^2$, with $\bar{\omega}_a=\omega_a(\eta+1)/2$ and $C=\omega_a(3\eta+1)(\eta-1)/8\eta(\eta+1)$. The corresponding positive eigenvalue represents the pseudo-isolated eigenfrequency in the AP,
\begin{equation}
\label{cs}
\tilde{\omega}_{0}= \frac{\omega_a}{2}\sqrt{\frac{(\eta-1)(\eta^2+6\eta+1)}
{\eta}}\, .
\end{equation}

\section{Heisenberg-Langevin Equations}
\label{sec:QLE}

We now study the open system in both phases with exact QLEs derived from the total Hamiltonian, \cref{eq:renormalized}, in the thermodynamic limit. Retaining the full counter-rotating structure of the coupling is essential once the system-bath rates become comparable to the low-energy mode frequency. Starting in the NP and tracing out the baths \cite{gardiner2004quantum}, the QLEs read, in the Fourier domain,
\begin{equation}
\label{M}
   i\,\mathbf{M}(\omega,\mathbf{A}_{NP},\pm\boldsymbol{\Gamma}_{NP})\,
   \mathbf{v}(\omega) = \boldsymbol{\tilde{F}}_{\mathrm{in/out}}(\omega)\, ,
\end{equation}
with $\mathbf{M}=\mathbf{A}_{NP}\mp i\boldsymbol{\Gamma}_{NP}/2-\omega \mathbf{I}$ (upper/lower sign for input/output), $\mathbf{v}=(\tilde a,\tilde a^{\dagger})^{T}$, Hopfield-Bogoliubov matrix $\mathbf{A}_{NP}$, decay matrix $\boldsymbol{\Gamma}_{NP}$, and Langevin force vectors ${\tilde F}_{\rm in/out}$ (explicit forms in SM).
In absence of coherent drive, i.e., $\langle{\tilde F}_{\rm in}\rangle=0$, the condition $\det\mathbf{M}(\omega,\mathbf{A}_{NP},\boldsymbol{\Gamma}_{NP})=0$ gives the complex eigenfrequencies in the NP ($\chi<\omega_a^2$),
\begin{equation}
\label{Omega}
  \Omega= -i\frac{\gamma_1+\gamma_2}{2}\pm\sqrt{\tilde{\omega}^2_0-\left(
  \frac{\gamma_1+\gamma_2}{2}\right)^2}\, .
\end{equation}
We identify the QPT critical point as the point where the complex excitation energy vanishes in both its real and imaginary parts; from \cref{Omega} this occurs at $\chi=\omega_a^2$, independent of $\gamma_1$ and $\gamma_2$ and equal to the pseudo-isolated value. The pseudo-isolated results follow from the QLEs by letting $\gamma_1,\gamma_2\to0$ at fixed $\chi\neq0$, as discussed above.

To describe the AP, we must include a possible macroscopic occupation of the ground state of the total (system-plus-reservoir) Hamiltonian \cite{DeLiberato2017}. Following Ref.~\cite{Lamberto2026superradiantquantum}, we shift both system and bath operators, $a=a_s+\alpha$ and $c_{jn}=c_{s,jn}+\sigma_{jn}$, and minimize the total Hamiltonian rather than the system part alone (see \cref{Minimization of the Total Hamiltonian in the Abnormal Phase}).
This reproduces the system condensate $\alpha^2$ of \cref{alpha} and additionally yields macroscopic bath occupations 
\begin{equation}
\label{sigma}
    \sigma_{jn}= \alpha\sqrt{\frac{\omega_{jn}}{\omega_a}}
    \sqrt{1-\frac{\alpha^2}{N}}\, .
\end{equation}
Notably, both reservoirs display a ground-state condensation, driven by the unstable bath. This feature is absent from frameworks that treat the baths perturbatively.

After this minimization, the resulting Hamiltonian for the AP can be written in a form analogous to \cref{eq:renormalized}, with shifted operators and renormalized parameters, and thus the QLE procedure can be applied as in the NP (see \cref{abnormal}). The obtained results reproduce the normal-phase structure, with a renormalization of the effective damping rates by a factor of $1/\eta$. This suppression originates from saturation effects intrinsic to the two-level nature of the spins. In particular, the ground-state condensate reduces, through Pauli blocking \cite{HEPP1973360,scully1997quantum,carmichael2013statistical}, the phase space for the transitions that produce decay, and it disappears continuously as $\chi\to \chi_c=\omega_a^2$ ($\alpha\to0$), restoring the full loss rate. Similarly to the NP, the resulting complex eigenfrequencies for $\chi>\omega_a^2$ are
\begin{equation}
\label{es}
\Omega=-i\frac{\gamma_1+\gamma_2}{2\eta} \pm\sqrt{\tilde{\omega}_0^2 -
\left(\frac{\gamma_1+\gamma_2}{2\eta} \right)^2}\, ,
\end{equation}
with $\tilde\omega_0$ from \cref{cs}.

\begin{figure}[t]
    \centering
    \includegraphics[width=\columnwidth]{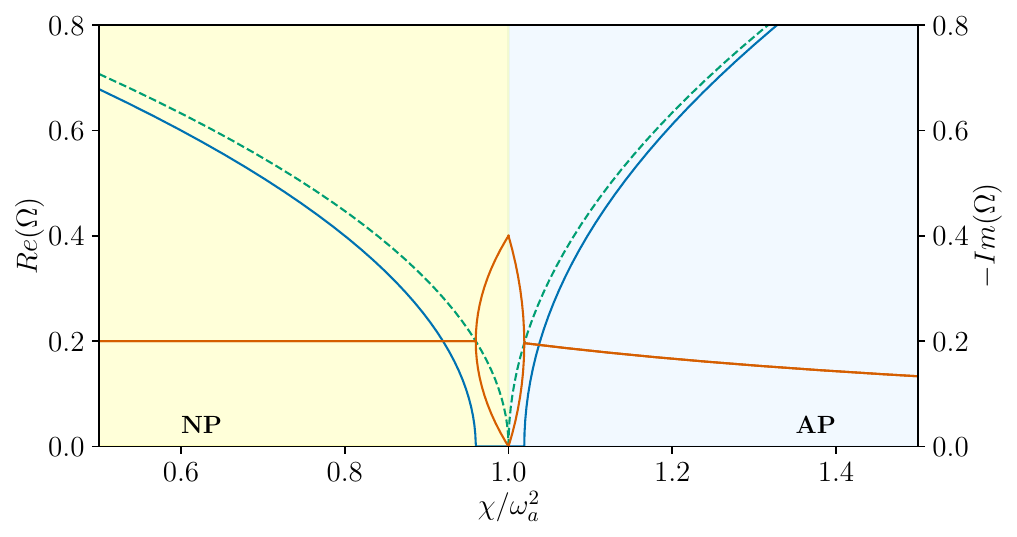}
    \caption{\textbf{Excitation Energies} Real (blue solid line) and negative imaginary (orange solid line) parts of the complex excitation energies as functions of the scaled integrated coupling strength $\chi/\omega_a^2$. The green dashed line shows the excitation energy of the pseudo-isolated system $\tilde\omega_0$. The yellow and light-blue shaded regions identify NP and AP, respectively. A finite gap region emerges around the critical point $\chi/\omega_a^2=1$, within which $\mathrm{Re}[\Omega]$ vanishes and $\mathrm{Im}[\Omega]$ splits into two branches. The vanishing of a complex excitation energy identifies the critical point. Parameters: $\omega_a=1$, $\gamma_1=0.3$ , $\gamma_1=0.1$}
    \label{fig:excitation}
\end{figure}

\begin{figure*}[t]
\centering
\includegraphics[width=\linewidth]{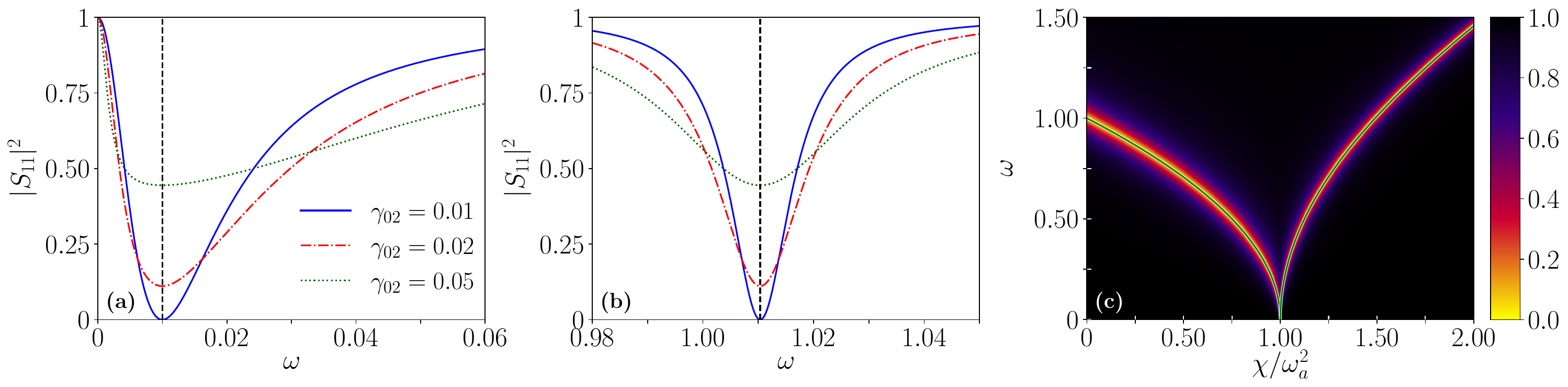}
\caption{\textbf{Reflection spectra.} (a) Normal phase, $\chi/\omega_a^2=0.9999$, and (b) abnormal phase, $\chi/\omega_a^2=1.5$, each for $\gamma_1=0.01\,\omega_a$ and $\gamma_2/\omega_a=0.01,0.02,0.05$ (solid, dash-dotted, and dotted curves, respectively).
(c) 2D map of $|S_{11}|^2$ as a function of $\chi/\omega_a^2$; the green line marks the closed-system frequency $\tilde{\omega}_0$.
Parameters in (c): $\gamma_1/\omega_a=\gamma_2/\omega_a=0.1$.}
\label{fig:spectra}
\end{figure*}
\section{Coherent Spectra}
\label{sec:spectra}

\Cref{fig:excitation} shows $\mathrm{Re}[\Omega]$ (blue) and $-\mathrm{Im}[\Omega]$ (orange) from \cref{Omega,es}, with the pseudo-isolated $\tilde\omega_0$ (green dashed). As $\chi$ increases at fixed $\gamma_{1,2}$, the soft mode carries the system through the three different regimes of a damped oscillator \cite{DEKKER19811}: \emph{underdamped} ($\tilde\omega_0>(\gamma_1+\gamma_2)/2$, $\mathrm{Re}[\Omega]\neq0$); \emph{critically damped} ($\tilde\omega_0=(\gamma_1+\gamma_2)/2$), where the modes coalesce into a single imaginary frequency; and \emph{overdamped} ($\tilde\omega_0<(\gamma_1+\gamma_2)/2$), where $\mathrm{Re}[\Omega]=0$ and $\mathrm{Im}[\Omega]$ splits into two branches.
This overdamped window is the finite gap separating NP and AP, which is absent in the pseudo-isolated case: it opens on the NP side once $\tilde\omega_0$ falls below $(\gamma_1+\gamma_2)/2$ and closes on the AP side once the abnormal soft mode rises back above $(\gamma_1+\gamma_2)/2\eta$. The critical point $\chi=\omega_a^2$ lies inside this gap, where one branch reaches $\Omega=0$, i.e., the simultaneous vanishing of real and imaginary parts that marks the transition \cite{PhysRevB.14.1165,PhysRevLett.90.227902,Osterloh2002}.
In the pseudo-isolated limit ($\gamma_i\to0$), the gap collapses to a point, recovering the sharp transition of the closed system \cite{PhysRevLett.101.105701,haken1975generalized}. Extending the normal-phase branch beyond $\chi=\omega_a^2$ would give the lower mode a positive imaginary part, i.e. a pole of the response functions (e.g., reflection and transmission coefficients) in the upper half of the complex plane, which violates causality.

We use input-output theory to relate the fields propagating to and from the system to its internal dynamics \cite{gardiner2004quantum,PhysRevA.29.1419,Savasta1996inputoutput}. Taking the continuum limit of the baths' degrees of freedom \cite{cohen2024photons}, we introduce input and output fields, $\tilde C_{{\rm in},j}(\omega)$ and $\tilde C_{{\rm out},j}(\omega)$ for each reservoir $j$, respectively, and define the scattering matrix (see \cref{Input-output fields and scattering matrix elements})

\begin{equation}
   S_{jk} = \left. \frac{\langle \tilde{C}_{\text{out},j} \rangle}
   {\langle \tilde{C}_{\text{in},k} \rangle} \right|_{\langle
   \tilde{C}_{\text{in},i} \rangle = 0,\ i \neq k}\, .
\end{equation}

As an example, the reflection coefficient at port 1 in the NP is given by
\begin{equation}
\label{S11NP}
S_{11}=\frac{\tilde{\omega}_0^2-\omega^2-i\omega[\gamma_2(\omega)-
\gamma_1(\omega)]}{\tilde{\omega}_0^2-\omega^2-i\omega[\gamma_2(\omega)+
\gamma_1(\omega)]}\, .
\end{equation}
Since the two reservoirs are the only channels through which energy leaves the system, the scattering matrix is unitary, so that the transmittance can be calculated through the relation $|S_{11}|^2+|S_{21}|^2=1$. 
In the AP,
\begin{equation}
\label{S11AP}
S_{11}(\omega)=\frac{\tilde{\omega}_0^2-\omega^2-i\omega
[\gamma_2(\omega)-\gamma_1(\omega)]/\eta}{\tilde{\omega}_0^2-\omega^2-
i\omega[\gamma_2(\omega)+\gamma_1(\omega)]/\eta}\, ,
\end{equation}
with $\tilde\omega_0$ now given by \cref{cs},
differing from the NP by the $1/\eta$ reduction of the rates due to the saturation effect discussed above. 

The minima of $|S_{11}|^2$ (both in NP and AP) are located at the pseudo-isolated frequency $\tilde\omega_0$, which near criticality differs from $\mathrm{Re}[\Omega]$ of the open system due to the presence of the imaginary part in \cref{Omega}. The depth of these minima is set by the rates alone, independent of proximity to criticality,
\begin{equation}
     |S_{11}(\tilde \omega_0)|^2 = \left( \frac{\gamma_2(\tilde \omega_0)-
     \gamma_1(\tilde \omega_0)}{\gamma_2(\tilde \omega_0)+\gamma_1(\tilde
     \omega_0)} \right)^2\, ,
\end{equation}
so that $\gamma_1(\tilde\omega_0)=\gamma_2(\tilde\omega_0)$ gives a reflectionless mode \cite{PhysRevA.102.063511,Bonizzoni2025,Rao2024}, corresponding to $|S_{11}|^2=0$. We stress that the framework is not restricted to constant (ohmic) rates and extends to arbitrary, frequency-dependent dissipation.

\Cref{fig:spectra}(a,b) shows $|S_{11}|^2$ near the critical point in the normal and abnormal phases, and (c) a 2D map as a function of $\chi/\omega_a^2$.
Even with non-negligible damping, the spectra track the closed-system frequency $\tilde\omega_0$, explaining why fits based on closed-system eigenfrequencies stay reliable in the critical region \cite{Kim2025}. The lineshape is markedly asymmetric \cite{Magazzù_2019}, broadened on the high-frequency side; the asymmetry sharpens toward criticality as the spectrum is compressed against the $\omega=0$ boundary, where the left wing of the resonance necessarily vanish.

\section{Conclusions}
\label{sec:concl}

We have shown that the very structure of a system-bath coupling can drive a QPT: an ensemble of two-level systems coupled to a counterterm-free bosonic reservoir undergoes mode softening and macroscopic condensation as the integrated coupling $\chi$ crosses $\omega_a^2$, with no coherent drive or additional light-matter coupling involved. 
An exact treatment based on the global minimization of the total system-bath Hamiltonian, and the subsequent use of QLEs, shows that this critical point coincides with that of the pseudo-isolated system and is independent of the dissipation rates, and that condensation develops simultaneously in the system and in both reservoirs. 
We also obtain closed-form reflection and transmission coefficients in both phases, whose mode softening and increasingly asymmetric lineshape are directly measurable signatures of the transition, even in the overdamped region around the critical point where Born-Markov and rotating-wave approximations break down.

These signatures are within reach in systems magnetically coupled to their environment, where a bilinear Zeeman-like interaction arises naturally without a counterterm. A yttrium iron garnet thin film on an open superconducting coplanar waveguide \cite{PhysRevA.69.062320} is a natural candidate: such films have recently reached the ultrastrong magnon-photon coupling regime with a counterterm-free interaction \cite{PhysRevApplied.20.024039,Han:21}, and replacing the resonator with an open waveguide would furnish exactly the dissipative, counterterm-free channel needed to observe \cite{Cai2021} the reservoir-induced softening of the lowest magnonic mode. 
The nominal critical condition $\chi\sim\omega_a^2$ may appear stringent, since the system-bath coupling typically lies below this threshold. However, it is substantially relaxed when the mechanism acts on top of a pre-existing Dicke-type softening \cite{PhysRevA.7.831,PhysRevE.67.066203,PhysRevLett.108.043003}, so that a moderate $\chi$ already suffices once the lower polariton has softened due to the light-matter interaction. More broadly, these results reinforce the view of dissipation not as a mere source of decoherence \cite{RevModPhys.89.035002,PhysRevLett.115.263601} but as an active ingredient capable of driving macroscopic QPTs \cite{Diehl2008,Verstraete2009}.
Capturing this regime, where losses dominate the softening mode, requires the exact, non-perturbative treatment of the environment as the one employed here.
Several directions follow naturally. The framework holds for arbitrary bath spectral densities, so the bath universality class (ohmic, sub- or super-ohmic) becomes a tunable handle on the critical behavior and its spectral fingerprints. The simultaneous condensation of both reservoirs suggests that the coherent bath occupation might itself be probed as an independent signature of the transition. Furthermore, the amplified fluctuations of the overdamped soft mode make the reservoir-induced critical point a promising resource for criticality-enhanced quantum sensing \cite{PhysRevLett.124.120504,PRXQuantum.6.020301,v7mf-yh8n}. 
Together, these open a route to harnessing quantum criticality through the deliberate design of system-bath interactions in open many-body systems.

\section{Acknowledgments}

S.S. acknowledge financial support from the National Recovery and Resilience Plan (PNRR) -- NextGenerationEU, Project Title “National Centre for HPC, Big Data and Quantum Computing (HPC)” -- Code National Center CN00000013 -- CUP D43C22001240001.

\section{Data availability}

All relevant data supporting the main conclusions of this article are available from the corresponding author upon reasonable request.

\appendix
\crefalias{section}{appendix}

\section{Quantum Langevin Equations in the Normal Phase}

Here, we derive the quantum Langevin equations (QLEs) governing the dynamics of the open system considered in the main text. The environment consists of two independent reservoirs described as infinite discrete sets of uncoupled harmonic oscillators coupled to the system through their coordinates. Specifically, we derive the general theoretical framework and apply it to the effective Hamiltonian in the normal phase; the abnormal phase is treated separately in \cref{abnormal}. Solving the equations of motion for the bath degrees of freedom in terms of the system variables and substituting back into the system's Heisenberg equations, we obtain the QLEs \cite{gardiner2004quantum}
\begin{widetext}
\begin{flalign}
\dot{Y}(t) = & \frac{i}{\hbar} [H_{\rm sys}, Y]
- \frac{i}{2\hbar}
\Bigg[
\xi_1(t)-f_{1}(t-t_0)X(t_0)
-\int_{t_0}^{t} f_1(t-t')\dot{X}(t')\,dt',
[X,Y]
\Bigg]_+ \notag\\
&-\frac{i}{\hbar}
\Bigg(
\xi_2(t)-f_2(t-t_0)X(t_0)+\chi X(t)
-\int_{t_0}^{t} f_2(t-t')\dot{X}(t')\,dt'
\Bigg)[X,Y] \quad ,
\end{flalign}
\end{widetext}
where $Y$ is an arbitrary system operator, $[...,...]_+$ is the anticommutator, $f_j(t)= \sum_{n}k_{jn}\cos{\omega_{jn} t} $ plays the role of a memory function (with $f_2(0)=\sum_{n}k_{2n}=\chi$ defined for convenience, see \cref{Memory_f_decay}), and
\begin{equation}
\label{rumore}
  \xi_{j}(t)=\sum_{n} k_{jn}  \sqrt{\frac{\hbar}{2\omega_{jn}}} \left[ e^{i \omega_{jn} (t-t_0)} c_{jn}^{\dagger}(t_0) + h.c. \right]
\end{equation}
are the baths noise terms. Therefore, the QLEs for the bosonic operators in the time domain are

\begin{align}\label{eq:bdot}
\dot{a}(t)=&-i\omega_a a(t)+\!\!\sum_{j=1,2}\!\frac{i}{\sqrt{2\hbar\omega_a}}
\left[\xi_j(t)-f_j(t\!-\!t_0)X_A(t_0)\right] \nonumber\\
&-\frac{i}{2\omega_a}\int_{t_0}^t f_j(t\!-\!t')
(\dot{a}(t')+\dot{a}^{\dagger}(t'))dt' \nonumber\\
&+\frac{i\chi}{2\omega_a}(a^{\dagger}(t)+a(t)) \,,
\end{align}
and the complex conjugate for $a^\dagger(t)$.

Now we examine   within the frequency domain, under the assumption that the initial condition was established far in the past, specifically as $t_0 \to -\infty$ . This assumption, together with the condition that $f(t)$ is sharply peaked around zero (for example a Dirac delta function in the case of an ohmic environment), allows us to neglect the terms involving $t-t_0$. So we can introduce the decay rate function $\gamma_j(\omega)$, defined as the Fourier transform of  $\theta(t)f_j(t)$, where $\theta(t)$ is the Heaviside step function, included to preserve causality. For the scope of this work, we treat $\gamma_j(\omega)$  as a real-valued function however it can be considered complex-valued in a more general framework. As a result, the QLEs describing the normal phase in the frequency domain are given by:
\begin{equation} \label{ch5:eq:QLE_inputapp}
-i\omega \begin{pmatrix}
\tilde{a}(\omega) \\
\tilde{a}^{\dagger}(\omega)
\end{pmatrix}
= -i\left(\mathbf{A}_{NP} - \frac{i}{2} \mathbf{\Gamma}_{NP}\right) \begin{pmatrix}
\tilde{a}(\omega) \\
\tilde {a}^{\dagger}(\omega)
\end{pmatrix}
\!+ \mathbf{\tilde{F}}_{\text{in}}(\omega)\,,
\end{equation}
where we defined the Langevin forces vector $\tilde{F}_{in}$ as
\begin{equation}
\label{force}
\tilde{F}_{in}(\omega) =
\begin{pmatrix}
\dfrac{i}{\sqrt{2 \hbar \omega_a}}\big(\tilde{\xi}_1(\omega)+\tilde{\xi}_2(\omega)\big) \vspace{0.15cm} \\
-\dfrac{i}{\sqrt{2 \hbar \omega_a}}\big(\tilde{\xi}_1(\omega)+\tilde{\xi}_2(\omega)\big)
\end{pmatrix} \, .
\end{equation}
The Hopfield-Bogoliubov and decay matrices, $\mathbf{A}_{\mathrm{NP}}$ and $\mathbf{\Gamma}_{NP}$, respectively, are expressed as
\begin{align}
    \mathbf{A}_{NP} &=
\begin{pmatrix}
\omega_a-\dfrac{\chi}{2\omega_a} & -\dfrac{\chi}{2\omega_a}  \\
\dfrac{\chi}{2\omega_a} & \dfrac{\chi}{2\omega_a}-\omega_a
\end{pmatrix} \, , \\
    \mathbf{\Gamma}_{NP} &= \left(\gamma_1(\omega)+\gamma_2(\omega)\right)
\begin{pmatrix}
1 
& -1 \\[6pt]
-1 
& 1
\end{pmatrix} \, .
\end{align}
The structure of $\mathbf{\Gamma}_{NP}$ is influenced by the specific modeling of system losses, which are fundamentally connected to the microscopic couplings between the system and the baths, $k_{jn}$ in Eq.~(2) in the main text. The off-diagonal elements of the matrix originate from counter-rotating contributions to the system-bath interaction.

\section{Relation between the Memory Function and the Bath Decay Rate}
\label{Memory_f_decay}

The memory function associated with the bath $2$ is defined, as in the previous section, as
\begin{equation}
    f_2(t)=\sum_n k_{2n}\cos(\omega_{2n}t) \,,
\end{equation}
where $k_{2n}$ represents the coupling strength between the system and the $2$-th environmental mode with frequency $\omega_{2n}$. The value of the memory function at the initial time, which we name $\chi$, is given by
\begin{equation} \label{app:eq_chi}
    \chi = f_2(0)= \sum_n k_{2n} \,\, ,
\end{equation}
The decay rate function $\gamma_2(\omega)$ is introduced as the Fourier transform of the causal memory kernel $\theta(t)f_2(t)$:
\begin{align}
    \gamma_2(\omega)&=
    \int_{-\infty}^{\infty} dt\, e^{i\omega t}\theta(t) f_2(t)=
    \int_0^{\infty} dt\, e^{i\omega t} f_2(t)= \nonumber \\ 
    &= \frac{1}{2}
\sum_n k_{2n}
\left[
\int_0^{\infty}dt\,e^{i(\omega+\omega_{2n})t}
+
\int_0^{\infty}dt\,e^{i(\omega-\omega_{2n})t}
\right] \, .
\end{align}
Using the Sokhotski-Plemelj relation
\begin{equation}
    \int_0^{\infty}dt\,e^{ixt}
    =
    \pi\delta(x)
    +
    i\mathcal{P}\left(\frac{1}{x}\right),
\end{equation}
where $\mathcal{P}$ denotes the Cauchy principal value, we obtain
\begin{align}
\gamma_2(\omega)&=\frac{\pi}{2}\sum_n k_{2n}\left[\delta(\omega-\omega_{2n})+\delta(\omega+\omega_{2n})\right] \nonumber\\
&\,\,+\frac{i}{2}\sum_n k_{2n}\left[\mathcal{P}\frac{1}{\omega-\omega_{2n}}+\mathcal{P}\frac{1}{\omega+\omega_{2n}}\right].
\end{align}
The first term represents the dissipative contribution, while the second one corresponds to the dispersive contribution associated with the frequency renormalization. 
In the following, we consider only the real part of the decay function.
For a physical bath composed of harmonic oscillators, the environmental frequencies are restricted to positive values, i.e. $\omega_{2n}>0$. Therefore, when considering the physical frequency domain $\omega>0$, the contribution associated with $\delta(\omega+\omega_{2n})$ vanishes, since its support is located at negative frequencies. Hence, the decay rate function reduces to
\begin{equation}
    \gamma_2(\omega)
    =
    \frac{\pi}{2}
    \sum_n k_{2n}
    \delta(\omega-\omega_{2n}),
    \qquad \omega>0 .
\end{equation}
Integrating this expression over the positive frequency domain gives
\begin{equation}
    \int_0^\infty d\omega\,\gamma_2(\omega)
    =
    \frac{\pi}{2}
    \sum_n k_{2n}
    \int_0^\infty d\omega\,
    \delta(\omega-\omega_{2n}) = \frac{\pi}{2}
    \sum_n k_{2n} \, .
\end{equation}
Therefore, from Eq.~\eqref{app:eq_chi}, we finally obtain the relation
\begin{equation}
    \chi=
    \frac{2}{\pi}
    \int_0^\infty d\omega\,\gamma_2(\omega) \, .
\end{equation}

\section{Minimization of the Total Hamiltonian in the Abnormal Phase}
\label{Minimization of the Total Hamiltonian in the Abnormal Phase}

In this section, we show a procedure to calculate the ground-state condensation in the abnormal phase. 
This method relies on minimizing the total system-bath Hamiltonian, providing a self-consistent framework that explicitly incorporates the bath's degrees of freedom.  
This constitutes a key difference from previous theoretical approaches, which usually assume weak system bath coupling and depend on the corresponding approximations. 
All calculations presented below are performed in the thermodynamic limit, retaining only states belonging to the maximum angular momentum manifold. This approximation is well justified: states with $j < N/2$ couple more weakly to the bosonic mode and therefore yield a negligible contribution to the emission spectra \cite{rgz2-4m69}. 

As already pointed out in the main text, to investigate the abnormal phase, we begin by shifting the bosonic operators of both the system and reservoirs as $a=a_s+\alpha$ , $c_{1n}=c_{s,1n}+\sigma_{1n}$ , $c_{2m}=c_{s,2m}+\sigma_{2m}$. The effective system coordinate in the abnormal phase to which the bath operators couple to, after expanding the square root as a Taylor series up to second order, is
\begin{equation}
    X_A=\sqrt{\hbar(N-\alpha^2)/2\omega_aN}\left(a_s^{\dagger}\sqrt{\xi}+\sqrt{\xi}a_s + 2\alpha\sqrt{\xi}\right) \, ,
\end{equation}
where $ \xi=1-[a_s^{\dagger}a_s+\alpha(a^{\dagger}_s+a_s)](N-\alpha^2)^{-1}$ . If we now substitute these definitions into the total Hamiltonian (Eq.~(2) of the main paper), and take into account the thermodynamic limit, the resulting expression can be written as (omitting constant contributions)
\begin{widetext}
\begin{align}
\label{Hamiltonian s}
\frac{H}{\hbar} &=\, \omega_a a_s^{\dagger} a_s 
+ \omega_a \alpha \left( a_s + a_s^{\dagger} \right)
+ \sum_{n} \Bigg\{
\frac{\alpha^2k_{1n}}{\omega_aN}a_s^{\dagger}a_s
+ \frac{k_{1n}}{4 \omega_{1n}} \left( c_{s,1n} + c_{s,1n}^{\dagger} \right)^2
- \frac{\omega_{1n}}{4} \left( c_{s,1n} - c_{s,1n}^{\dagger} \right)^2 \notag \\
&\,\,
+ \!\!\left( \frac{k_{1n} \sigma_{1n}}{\omega_{1n}}
-\frac{\alpha k_{1n}}{\sqrt{\omega_{1n}\omega_a}}
\sqrt{\frac{N-\alpha^2}{N}} \right) \!\!
\left( c_{s,1n} + c_{s,1n}^{\dagger} \right)
\!\!+\!\! \left(
\frac{k_{1n}\alpha^2}{2\sqrt{N\omega_{1n}\omega_a(N-\alpha^2)}}
- \frac{k_{1n}}{2}
\sqrt{\frac{N - \alpha^2}{N \omega_{1n} \omega_a}}
\right)\!\!
\left( c_{s,1n} + c_{s,1n}^{\dagger} \right)\!\!
\left( a_s + a_s^{\dagger} \right) \notag \\
&\,\,
+\!\! \left(
\frac{k_{1n}(\alpha N - 2\alpha^3)}{\omega_a N}
- \frac{k_{1n}\alpha(N-\alpha^2)}{\omega_aN}
+\frac{\alpha^3 k_{1n}}{\omega_aN}
\right)\!\!
\left(a_s + a_s^{\dagger}\right)
+ \left(
\frac{k_{1n}(N - 5\alpha^2)}{4 \omega_a N}
+ \frac{\alpha^4 k_{1n}}{4 \omega_a N (N - \alpha^2)}
+\frac{\alpha^2 k_{1n}}{2\omega_aN}
\right)\!\!
\left( a_s + a_s^{\dagger} \right)^2 \!
\Bigg\} \notag \\
&\,\,
+ \sum_{m} \Bigg\{
\frac{k_{2m}}{4 \omega_{2m}}
\left( c_{s,2m} + c_{s,2m}^{\dagger} \right)^2
- \frac{\omega_{2m}}{4}
\left( c_{s,2m} - c_{s,2m}^{\dagger} \right)^2
+ \left(
\frac{k_{2m} \sigma_{2m}}{\omega_{2m}}
-\frac{\alpha k_{2m}}{\sqrt{\omega_{2m}\omega_a}}
\sqrt{\frac{N-\alpha^2}{N}}
\right)
\left( c_{s,2m} + c_{s,2m}^{\dagger} \right) \notag \\
&\,\,
+ \left(
\frac{k_{2m}\alpha^2}{2\sqrt{N\omega_{2m}\omega_a(N-\alpha^2)}}
- \frac{k_{2m}}{2}
\sqrt{\frac{N - \alpha^2}{N \omega_{2m} \omega_a}}
\right)
\left( c_{s,2m} + c_{s,2m}^{\dagger} \right)
\left( a_s + a_s^{\dagger} \right)
+ \left(
\frac{\alpha^3k_{2m}}{\omega_aN}
-\frac{k_{2m}\alpha(N-\alpha^2)}{\omega_aN}
\right)
(a_s+a_s^{\dagger}) \notag \\
&\,\,
+ \frac{\alpha^2 k_{2m}}{\omega_aN}a_s^{\dagger}a_s
+ \left(
\frac{\alpha^4 k_{2m}}{4 \omega_a N (N - \alpha^2)}
+\frac{\alpha^2 k_{2m}}{2\omega_aN}
\right)
\left( a_s + a_s^{\dagger} \right)^2
\Bigg\} \,\, .
\end{align}
\end{widetext}
We now proceed to minimize the Hamiltonian by requiring the cancellation of the linear terms in the bosonic operators appearing in \cref{Hamiltonian s}, leading to the non-zero macroscopic occupation numbers
\begin{equation}
    \alpha^2=\frac{N}{2}\left(1-\frac{1}{\eta}\right) ,\quad \sigma_{jn}= \alpha\sqrt{\frac{\omega_{jn}}{\omega_a}}\sqrt{1-\frac{\alpha^2}{N}} \quad ,
\end{equation}
where $\eta=\chi / \omega_a^2$ as in the main text. Inserting these equilibrium values into \cref{Hamiltonian s}, the effective total Hamiltonian for the abnormal phase simplifies to
\begin{align}
\label{supera}
H=&\,\bar{\omega}_a a_s^{\dagger}a_s+C(a_s^{\dagger}+a_s)^2 \nonumber\\
&+\sum_n\left(\frac{p_{s,1n}^2}{2}+\frac{k_{1n}}{2}q_{s,1n}^2-k_{1n}q_{s,1n}x_a+\frac{k_{1n}}{2}x_a^2\right) \nonumber\\
&+\sum_m\left(\frac{p_{s,2m}^2}{2}+\frac{k_{2m}}{2}q_{s,2m}^2-k_{2m}q_{s,2m}x_a\right).
\end{align}
where $x_a=(a_s+a_s^{\dagger})/\sqrt{\omega_a\eta(\eta+1)}$, while $q_{s,jn}$ and $p_{s,jn}$ are the shifted baths' coordinates and momenta.

\section{Quantum Langevin Equations in the abnormal phase}
\label{abnormal}
To derive the QLEs in the abnormal phase, we follow the same methodology as in the normal phase. Indeed, the Hamiltonians for the normal and abnormal phases share the same formal structure, the only difference being that the bare bosonic operators are replaced by their shifted counterparts and the parameters are suitably renormalized. The QLEs in the abnormal phase thus read
\begin{equation}
\begin{aligned}
\dot{a}_s(t)=&-i\bar{\omega}_a a_s(t)-2iC\big(a_s(t)+a_s^{\dagger}(t)\big)\\
&+\sum_{j=1,2}\frac{i}{\sqrt{\omega_a\eta(\eta+1)}}\xi_j(t)\\
&+\frac{1}{2\eta}\int_{-\infty}^{\infty}f_j(t-t')\theta(t-t')
\big(a_s^{\dagger}(t)-a_s(t)\big)\,dt'\\
&+\frac{i\chi}{\omega_a\eta(\eta+1)}
\big(a_s^{\dagger}(t)+a_s(t)\big) \, ,
\end{aligned}
\end{equation}
where $\bar{\omega}_a$, $\eta$ and $C$ are defined in the main text. An analogous equation holds for $\dot{a}_s^\dagger(t)$. These equations can be reformulated within the frequency domain as
\begin{equation}
\label{superadmatrix}
-i\omega 
\textbf{v}(\omega)
= 
-i\left(\mathbf{A}_{AP}(\omega) - \frac{i}{2}\mathbf{\Gamma}_{AP}(\omega)\right) \!\!
 \textbf{v}(\omega)
\!+ \tilde{F}_{\text{in},a}(\omega) \, ,
\end{equation}
where we defined the vector $\mathbf{v}(\omega) = \left(\tilde{a_s}(\omega), \tilde{a_s}^\dagger(\omega) \right)^T$  and the Langevin forces vector $\tilde{F}_{in,a}(\omega)$ in the abnormal phase as
\begin{equation}
\tilde{F}_{in,a}(\omega) =
\begin{pmatrix}
\dfrac{i}{\sqrt{\hbar\omega_a\eta(\eta+1)}}\big(\tilde{\xi}_1(\omega)+\tilde{\xi}_2(\omega)\big) \vspace{0.15cm} \\
-\dfrac{i}{\sqrt{\hbar\omega_a\eta(\eta+1)}}\big(\tilde{\xi}_1(\omega)+\tilde{\xi}_2(\omega)\big)
\end{pmatrix} \,\,.
\end{equation}
By introducing the matrix $\mathbf{M}(\mathbf{A}_{AP},\mathbf{\Gamma}_{AP})= \mathbf{A}_{AP}-\frac{i}{2}\mathbf{\Gamma}_{AP}-\omega \mathbf{I}$, \cref{superadmatrix} can be written in a more compact form as 
\begin{equation}
    i\mathbf{M}(\mathbf{A}_{AP},\mathbf{\Gamma}_{AP})\mathbf{v}= \mathbf{\tilde{F}}_{in,a} \, .
\end{equation}
For the sake of convenience, we have simplified the notation, and removed the explicit dependence on $\omega$. For completeness, we note that an analogous equation applies to the output fields, namely
\begin{equation}
    i\mathbf{M}(\mathbf{A}_{AP},-\mathbf{\Gamma}_{AP})\mathbf{v}= \mathbf{\tilde{F}}_{out,a}\, .
\end{equation}
As in the normal phase, $\mathbf{A}_{AP}$ and $\mathbf{\Gamma}_{AP}$ denote the Hopfield-Bogoliubov and decay matrices in the abnormal phase, respectively, given by 
\begin{align}
\renewcommand{\arraystretch}{1.2}
\mathbf{A}_{AP}(\omega) &=
\begin{pmatrix}
\bar{\omega_a}+2C - \dfrac{\chi}{2\bar{\omega}_a\eta}  & 2C - \dfrac{\chi}{2\bar{\omega}_a\eta} \vspace{0.15cm} \\
 -2C + \dfrac{\chi}{2\bar{\omega}_a\eta}& -\bar{\omega_a}-2C+ \dfrac{\chi}{2\bar{\omega}_a\eta}
\end{pmatrix} \, , \\
\normalsize
    \mathbf{\Gamma}_{AP} &= (\gamma_1(\omega)+\gamma_2(\omega))
\begin{pmatrix}
1/\eta
& -1/\eta \\[6pt]
-1/\eta 
& 1/\eta
\end{pmatrix} \, .
\end{align}
The prefactor $1/\eta$ is linked to a saturation effect intrinsic to the two-level nature of the constituent spins, as discussed in the main text: in the abnormal phase, the ground state acquires a macroscopic occupation, and since the decay rate describes transitions from the excited states to the ground state, this occupation reduces the phase space available for such transitions. Therefore, the presence of the condensate suppresses the effective decay rate by a factor of $1/\eta$.

\section{Input-output fields and scattering matrix elements}
\label{Input-output fields and scattering matrix elements}
A precise definition of input and output fields is crucial for enabling the analytical computation of reflection and transmission spectra. Following the approach outlined in ref.~\cite{gardiner2004quantum}, we define these fields in terms of initial and final conditions. In particular, $t_0<t$ denotes an initial time, typically assumed to lie in the remote past, while $t_1>t$ corresponds to a final time, generally considered to be in the distant future. These fields read:

\begin{equation}
\label{input}
\begin{aligned}
C_{in,j}(t) = \frac{1}{2}\int_{0}^{\infty} \sqrt{\frac{\hbar}{\pi \omega}} & \left( \tilde{c}_{in,j}(\omega,t_0)e^{-i\omega( t-t_0)} \right. \\
& \left. + \tilde{c}_{in,j}^{\dagger}(\omega,t_0)e^{i\omega (t-t_0)} \right) d\omega,
\end{aligned}
\end{equation}

\begin{equation}
\label{output}
\begin{aligned}
C_{out,j}(t) = \frac{1}{2}\int_{0}^{\infty} \sqrt{\frac{\hbar}{\pi \omega}} & \left( \tilde{c}_{out,j}(\omega,t_1)e^{-i\omega( t-t_1)} \right. \\
& \left. + \tilde{c}_{out,j}^{\dagger}(\omega,t_1)e^{i\omega (t-t_1)} \right) d\omega,
\end{aligned}
\end{equation}

where $\tilde{c}_{in,j}(\omega)=\tilde{c}_j(\omega,t_0)e^{i\omega t_0}$ is the $j$-th bath bosonic operator evaluated in the remote past, ideally $t_0 \to -\infty$ . The corresponding output operators, $\tilde{c}_{out,j}(\omega)=\tilde{c}_j(\omega,t_1)e^{i\omega t_1}$, are defined similarly with the only distinction that the bath operators are evaluated in the distant future, $t_1 \to \infty$ . These operators obey the standard canonical commutation rules
\begin{equation}
\left[\tilde{c}_{in/out,j}(\omega'),\tilde{c}_{in/out,j}^{\dagger}(\omega'')\right]=\delta_{jk}\delta(\omega'-\omega'') \, .
\end{equation}
To simplify the analysis, we applied the continuum limit to the bath's degrees of freedom. It is important to note that when the rotating wave approximation is not used, both co-rotating and counter-rotating components in $C_{in,j}(t)$ and $C_{out,j}(t)$ must be taken into account. We are now ready to establish the connection between the Langevin forces and the input/output fields. This begins with applying the continuum limit to $\xi_j(t)$, which is expressed as follows:
\begin{equation}
\label{re}
\xi_{j}(t)= \int_{0}^{\infty} \sqrt{\frac{\hbar \omega k_j(\omega)}{2}}\left(\tilde{c}_{j}(\omega,t_0)e^{-i\omega (t-t_0)}+\textbf{h.c}\right) d\omega \, .
\end{equation}
By using the definition of the damping rate, it can be shown that $k_j(\omega)=2\gamma_j(\omega)/\pi$, so that \cref{re} closely resembles the input operators described in \cref{input}, which leads to the corresponding expression in the frequency domain
\begin{equation}
\label{rumore}
   \tilde{\xi}_j(\omega) \propto\omega \sqrt{\gamma_j(\omega)}\tilde{C}_{in,j}(\omega)
\end{equation}
It is now straightforward to establish the connection between the Langevin forces and the input/output operators in Fourier space
\begin{equation}
\label{campi}
    \tilde{C}_{in/out,j} \propto \pm \sqrt{\frac{\omega_a}{\gamma_j}} \frac{\tilde{F}_{in/out,j}}{\omega}
\end{equation}
where the $+ (-)$ is associated to the annihilation (creation) system operator. We now derive the input-output relation \cite{Lamberto2026superradiantquantum}, linking the input and output operators to the system operators. To this end, we write the QLEs for the annihilation operator $\tilde{a}(\omega)$ in the frequency domain, first expressed in terms of the input fields and then in terms of the output fields:
\begin{subequations}
\label{eq:moto_tilde}
\begin{align}
-i\omega \tilde{a}(\omega) =\;& -i\omega_a \tilde{a}(\omega) + \frac{i\omega\sqrt{\gamma_1}}{\sqrt{2\omega_a}} \tilde{C}_{in,1}(\omega) + \frac{i\omega\sqrt{\gamma_2}}{\sqrt{2\omega_a}} \tilde{C}_{in,2} (\omega)\nonumber \\
&+ \frac{\gamma_1 + \gamma_2}{2}(\tilde{a}^\dagger(\omega) - \tilde{a}(\omega)) + \frac{i\chi}{2\omega_a}(\tilde{a}^\dagger(\omega)+ \tilde{a}(\omega)) \, ,
\label{eq:moto_tilde_a} \\[10pt]
-i\omega \tilde{a}(\omega) =\;& -i\omega_a \tilde{a}(\omega) + \frac{i\omega\sqrt{\gamma_1}}{\sqrt{2\omega_a}} \tilde{C}_{out,1}(\omega) + \frac{i\omega\sqrt{\gamma_2}}{\sqrt{2\omega_a}} \tilde{C}_{out,2}(\omega) \nonumber \\
&- \frac{\gamma_1 + \gamma_2}{2}(\tilde{a}^\dagger(\omega) - \tilde{a}(\omega)) + \frac{i\chi}{2\omega_a}(\tilde{a}^\dagger(\omega) + \tilde{a}(\omega)) \, .
\label{eq:moto_tilde_b}
\end{align}
\end{subequations}
Subtracting these two equations, we obtain the input-output relation
\begin{equation}
\label{eq:io}
\tilde{C}_{in,j}(\omega)=\tilde{C}_{out,j}(\omega)+\frac{\sqrt{2\gamma_j(\omega)\omega_a}}{i\omega}(\tilde{a}(\omega)-\tilde{a}^\dagger(\omega)) \, .
\end{equation}

To derive $S_{11}$ and $S_{21}$, we explicitly solve the QLEs for $\tilde{a}(\omega)$ and $\tilde{a}^{\dagger}(\omega)$ in terms of the input fields:

\begin{subequations}
\label{eq:moto_tilde}
\begin{align}
-i\omega \tilde{a}(\omega) = \;& -i\omega_a \tilde{a}(\omega) + \frac{i\omega\sqrt{\gamma_1(\omega)}}{\sqrt{2\omega_a}} \tilde{C}_{in,1}(\omega) \nonumber \\
& + \frac{i\omega\sqrt{\gamma_2(\omega)}}{\sqrt{2\omega_a}} \tilde{C}_{in,2}(\omega) + \frac{i\chi}{2\omega_a}(\tilde{a}^\dagger(\omega) + \tilde{a}(\omega)) \nonumber \\
& + \frac{\gamma_1(\omega) + \gamma_2(\omega)}{2}(\tilde{a}^\dagger(\omega) - \tilde{a}(\omega)) \, , \label{eq:moto_tildeb} \\[12pt]
-i\omega \tilde{a}^{\dagger}(\omega) = \;& i\omega_a \tilde{a}^{\dagger}(\omega) -\frac{i\omega\sqrt{\gamma_1(\omega)}}{\sqrt{2\omega_a}} \tilde{C}_{in,1}(\omega) \nonumber \\
& - \frac{i\omega\sqrt{\gamma_2(\omega)}}{\sqrt{2\omega_a}} \tilde{C}_{in,2}(\omega) - \frac{i\chi}{2\omega_a}(\tilde{a}^\dagger(\omega) + \tilde{a}(\omega)) \nonumber \\
& - \frac{\gamma_1(\omega) + \gamma_2(\omega)}{2}(\tilde{a}^\dagger(\omega) - \tilde{a}(\omega)) \, . \label{eq:moto_tilde_bdagger}
\end{align}
\end{subequations}

By algebraically solving this linear system for $\tilde{a}(\omega)$ and
$\tilde{a}^\dagger(\omega)$, and substituting the result back into the
input-output relation \cref{eq:io}, we can express the output fields entirely in terms of the input fields. Considering no incoming signal at port~2, i.e.\ $\langle\tilde{C}_{in,2}(\omega)\rangle=0$, we finally obtain the scattering matrix elements
\begin{equation}
 S_{11}=\frac{\tilde{\omega}_0^2-\omega^2-i\omega(\gamma_2(\omega)-\gamma_1(\omega))}{\tilde{\omega}_0^2-\omega^2-i\omega(\gamma_2(\omega)+\gamma_1(\omega))}  \, ,
 \end{equation}
 \begin{equation}
S_{21}=\frac{2i\omega\sqrt{\gamma_1(\omega)\gamma_2(\omega)}}{\tilde{\omega}_0^2-\omega^2-i\omega(\gamma_2(\omega)+\gamma_1(\omega))} \, .
\end{equation}
An analogous procedure can be applied in the abnormal phase, where the
additional prefactor $\sqrt{2/\eta(\eta+1)}$ introduced yields modified expressions for $S_{11}$ and $S_{21}$ that account the emergence of condensates in the system's ground state, i.e.,
\begin{equation}
 S_{11}=\frac{\tilde{\omega}_0^2-\omega^2-\frac{i\omega}{\eta}(\gamma_2(\omega)-\gamma_1(\omega))}{\tilde{\omega}_0^2-\omega^2-\frac{i\omega}{\eta}(\gamma_2(\omega)+\gamma_1(\omega))} \, , 
\end{equation}
\begin{equation}
S_{21}=\frac{\frac{2i\omega}{\eta}\sqrt{\gamma_1(\omega)\gamma_2(\omega)}}{\tilde{\omega}_0^2-\omega^2-\frac{i\omega}{\eta}(\gamma_2(\omega)+\gamma_1(\omega))} \, .
\end{equation}

\bibliography{bibliografia}

\end{document}